\documentclass{elsart}
\usepackage{amssymb}

\begin{document}

\begin{frontmatter}

\title{Superconducting protons in metals}

\author{J.S.Brown}
\ead{j.brown3@physics.ox.ac.uk}
\address{University of Oxford, Clarendon Laboratory, Parks Rd., OX1 3PU, England}

\begin{abstract}
The hitherto neglected phonon-exchange interaction between interstitial protons in metal lattices is found to be large. It is shown that this effect may give rise to a phase of protonic superconductivity, characterized by the formation of Cooper-like pairs of protons, in certain metals at high stoichiometric loading.
\end{abstract}

\end{frontmatter}

\section{Introduction}
The work reported in this Letter was motivated by the observation that, despite an extensive theoretical literature \cite{Sug80,Pus83,Pus84,Fis89,Tom91,Kar00,Sun04,Bar04} dealing with what are often quite subtle effects in hydrogenated metals, no qualitative treatment of the phonon-exchange interaction between protons has ever been presented. It is hoped that the approximate model outlined here goes some way towards remedying this situation.
\\
The matrix element in the Fr\"ohlich model for the (energy-conserving) exchange of a phonon between two charged particles in momentum states $\{\vec k, -\vec k\}$ is (omitting spin indices)
\begin{equation}
\label{hkwq}
H^{ph}_k = \sum_q |M_q|^2 \biggl [ \frac{\omega_q}{(E_{k+q} - E_k)^2 - \omega_q^2} \biggr ] c^{\dag}_{k+q}  c_k c^{\dag}_{-k-q} c_{-k}
\end{equation}
where $\omega_q \approx \omega_D \sin(\vec q \cdot \vec a)$.
\\
It has been known for a long time\cite{Mig58} that higher-order vertices contribute terms of the order of $(m/M_{ion})^{1/2}$ and smaller, so (\ref{hkwq}), although normally considered to apply only to electrons, should also provide a reasonable description of the interaction of interstitial protons in metal lattices.
\\
At the electron densities prevailing in metals, the screening of a proton is well-described by the nonlinear response theory \cite{Pop74,Alm76}. For simplicity, we approximate the instantaneous potential due to a proton in a lattice of cell dimension $a$ by
\begin{equation}
\label{yuk}
V(r) \approx \biggl (\frac{e}{r} \biggr) \cdot \mathrm{e}^{- k_{NL} r} \hspace{10mm}  \mathrm{where} \hspace{5mm} \ k_{NL} \approx 2 / a
\end{equation}
Consistency with the classical model of the dielectric effect of the slow moving lattice ions convoluted with the adiabatic response of the electron gas requires 
\begin{equation}
|M_q|^2 = \frac{Z^2 e^2 \omega_q}{V(q^2 + k_{NL}^2)}
\end{equation}
The instantaneous and retarded interactions will be subject to exactly the same electronic screening response. The total net interaction between a pair of interstitial particles of effective charge $Z$ and mass $m$ is accordingly
\begin{equation}
\label{hkq}
H_k^I = \sum_q \frac{Z^2 e^2}{q^2 + k^2_{NL}}\biggl[\frac{(q^2 - 2 \vec k \cdot \vec q)^2}{(q^2 - 2 \vec k \cdot \vec q)^2 - m^2 \omega_D^2 \sin^2(\vec q \cdot \vec a)} \biggr]c^{\dag}_{k+q} c_k c^{\dag}_{-k-q} c_{-k}
\end{equation}

This term is the basis of the BCS theory of low temperature superconductivity. There is a qualitative difference from the electronic case however. Whereas for electrons of effective mass $m^*_e$ the phonon energies $0 < \omega_q < \omega_D$ 
are typically smaller than $\Delta E = E_{k+q} - E_k$ everywhere except near the Fermi surface, the phonon and
quasi-free proton bandwidths are very comparable for a wide range of metals because $a^2 m_p\omega_D \approx 1$. 
There are thus strong {\em a priori} grounds for suspecting the existence of paired proton states having binding 
energies {\em considerably} greater than those typical for Cooper pairs, in a wide range of crystalline environments. The thermodynamic preconditions necessary for the actual occurrence of such states will be investigated further in Section \ref{mean} below.
\section{Free particle limit}
In order to get a rough idea of the properties that such hypothetical pairs might be expected to exhibit, it is necessary to solve (\ref{hkq}) explicitly. In anticipation of the large binding energy, we neglect the interaction with the lattice. Even with this simplification, the two-particle Hamiltonian is still not amenable to analytic solution owing to the k-dependence of the interaction potential and the periodicity of the phonon dispersion curve. However, if one considers only l=0 states, the two-body system can be transformed into a central force problem with spherical Bessel basis set
\begin{equation}
\varphi(r) = \sum_{n=1}^N c_n \frac {\sin (2\pi n r /R)}{r \sqrt{R}}
\end{equation}
The resultant k-space matrix can be solved by computer to an accuracy limited only by the value of the cut-off momentum $N/R$. 
\\
With a matrix dimension of N=1000, the calculated energy values and state vectors have been found to vary smoothly over a wide range of physically realistic values of all the relevant parameters. The dependencies can be summarized by the following simple generic formulae for the ground state
\begin{equation}
\label{e0}
E_0 \approx - 4 Ze \cdot m \biggl [ \frac{\omega_D}{k_{NL}} - \frac{4 \cdot 10^{-6}}{ \sqrt{k_{NL}R} } \biggr ] \hspace{10mm} \mathrm{and} \hspace{5mm} 
\langle r \rangle \approx \frac{2}{k_{NL}}
\end{equation}

Accordingly, given the assumptions of our model, we conclude that:

i) the ground state binding energy is of the order of 1 eV and therefore dominates over the periodic lattice potential, as expected
\\
ii) the binding energy is proportional to nucleon mass $m$, (subject to $m \ll M$), the bare nucleon charge $Ze$ and the Debye frequency $\omega_D$
\\
iii) there exists a certain critical crystal size of the order of a few nm below which no bound state nucleon pairs can form.
\\
iv) the radial extent of the pair is largely determined by the adiabatic screening distance and is therefore orders of magnitude smaller than that typical for an electronic Cooper pair. 
\\
Our numerical calculations also revealed excited bound states, provided the crystal is large enough. The spectrum is reminiscent of the Rydberg series for hydrogen.

\section{Mean-field Hubbard model}
\label{mean}
In his analysis of the thermodynamic phases of narrow half-filled bands, Bari
has shown \cite{Bari73} that (\ref{hkq}) admits of a particularly simply approximate form in the Wannier representation when $a^2 m\omega_D$ is large. Following the approach used in that earlier work, we confirm that for a cubic lattice of $N$ interstitial sites
\begin{displaymath}
H_{i,i',j,j'}^I = N^{-2} \sum_k H_k^I \mathrm{e}^{-\mathrm{i} \bigl[ \vec k \cdot (\vec R_i - \vec R_{i'} - \vec R_j + \vec R_{j'}) + \vec q \cdot (\vec R_{j'} - \vec R_j)  \bigr] } c^{\dag}_j c^{\dag}_{j'} c_i c_{i'}
\end{displaymath}
\begin{equation}
\approx I_1 c^{\dag}_i c^{\dag}_i c_i c_i -\frac{I_2}{N} c^{\dag}_j c^{\dag}_j c_i c_i
\end{equation}
where $c_i$ is the operator removing a proton at site $i$, we have omitted the spin indices and $I_1$ is the dominant on-site term in (\ref{yuk}). In arriving at this grossly simplified form, we have used the fact that the major contributions come from $\{k,q\} > a^{-1}$ and therefore the $\{i,j\} = \{i',j'\}$ terms dominate. 
\\
In view of (\ref{e0}), $I_2 \approx 1eV$. The hopping term $t_{ij}c^{\dag}_i c_j$ in hydrogenated metals \cite{Sun04,Sey75} is tiny compared to $I_1$ and $I_2$ and can be ignored. The effective mean-field Hamiltonian for the $i$th interstitial site is accordingly given by
\begin{equation}
H_{eff} = I_1 c^{\dag}_i c^{\dag}_i c_i c_i - I_2( \alpha c^\dag_i c^\dag_i + \alpha^* c_i c_i ) - \mu c^{\dag}_i c_i
\end{equation}
where the off-diagonal order parameter for the superconducting pair state $\alpha = \langle c_i c_i \rangle$ must be determined by iteration to self-consistency. 
\\
$\mu$ is a chemical potential for setting the mean site occupancy $\langle n \rangle$.
\\
In view of the identities $\langle c_{i\uparrow} c_{i\uparrow} \rangle = \langle c_{i\downarrow} c_{i\downarrow} \rangle = 0$, there are just 4 possible Wannier site states, with explicit Hamiltonian representation
\begin{equation}
\mathbf{H}| \varphi \rangle = \left ( \begin{array}{cccc} 0 & 0 & 0 & -I_2 \alpha^* \\ 0 & -\mu & 0 & 0 \\ 0 & 0 & -\mu & 0 \\ - I_2 \alpha & 0 & 0 & I_1 - 2\mu \end{array} \right )
\left ( \begin{array}{cccc} | \emptyset \rangle \\ | \uparrow \rangle \\ | \downarrow \rangle \\ |\uparrow \downarrow \rangle \end{array} \right )
\end{equation}
The one-particle partition function, site energies and mean occupancy are
\begin{equation}
Z = 2\mathrm{e}^{ \beta \mu} + \mathrm{e}^{-\beta E_-} + \mathrm{e}^{-\beta E_+}
\end{equation}

\begin{equation}
2E_{\pm} = I_1 - 2\mu \pm \bigl [ (I_1 - 2\mu)^2 + 4(I_2)^2 |\alpha|^2 \bigr ]^{1/2}
\end{equation}
\begin{equation} 
\langle n \rangle = \frac{1}{Z} \biggl [ \mathrm{e}^{ \beta \mu } + \frac{E_+^2 \mathrm{e}^{-\beta E_+}}{E_+^2 + (I_2)^2 \alpha^2}  + \frac{E_-^2  \mathrm{e}^{-\beta E_-}}{E_-^2 + (I_2)^2 \alpha^2} \biggr ]
\end{equation}
\begin{equation}
\alpha := \hspace{10mm} Z = \frac{I_2 \alpha E_+ \mathrm{e}^{-\beta E_+}}{ E_+^2 + (I_2)^2 \alpha^2}  + \frac{I_2 \alpha E_- \mathrm{e}^{-\beta E_-}}{E_-^2 + (I_2)^2 \alpha^2} 
\end{equation}
Clearly, $\beta I_1 \gg 1$ at $300 K$. In this "low-T" regime, provided $I_2 \gtrapprox 2I_1$, there is a transition from the pure interstitial state characterized by $\alpha=0$ to a mixed phase with $\alpha > 0$ above a certain  value of $\langle n \rangle < 1$. The sharpness of this transition increases with $\lambda \equiv I_2/I_1$ as shown below:
\\
\vspace{5mm}
\setlength{\unitlength}{5cm}
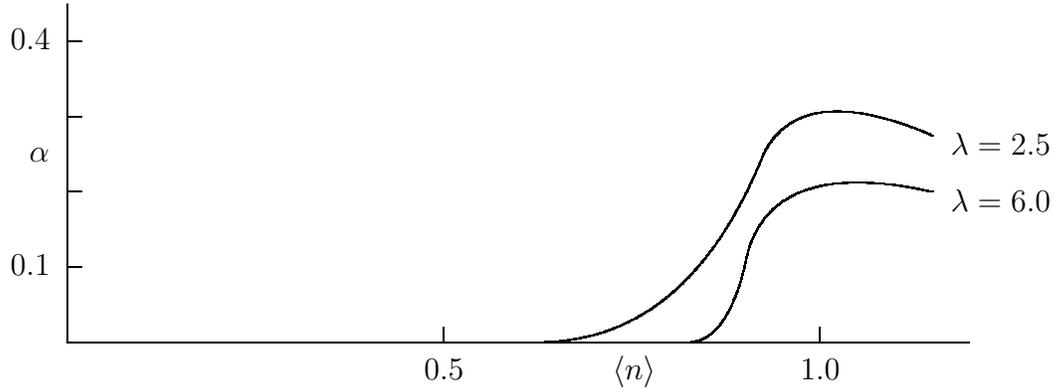
\begin{picture}
(2.5,1.2)
\put(0.15,0.1){\line(0,1){0.9}}

\put(0.15,0.3){\line(1,0){0.04}}
\put(0.0,0.28){$0.1$}
\put(0.15,0.5){\line(1,0){0.04}}
\put(0.05,0.58){$\alpha$}
\put(0.15,0.7){\line(1,0){0.04}}
\put(0.15,0.9){\line(1,0){0.04}}
\put(0.0,0.88){$0.4$}
\put(1.15,0.1){\line(0,1){0.04}}
\put(1.10,0){$0.5$}
\put(1.60,0){$\langle n \rangle$}
\put(2.15,0.1){\line(0,1){0.04}}
\put(2.10,0){$1.0$}
\put(0.15,0.1){\line(1,0){2.4}}
\qbezier(1.40,0.1)(1.8,0.1)(2.0,0.6)
\qbezier(2.0,0.6)(2.1,0.8)(2.45,0.65)
\put(2.50,0.6){$\lambda = 2.5$}

\qbezier(1.80,0.1)(1.90,0.1)(1.95,0.3)
\qbezier(1.95,0.3)(2.0,0.6)(2.45,0.5)
\put(2.50,0.45){$\lambda = 6.0$}
\end{picture}
\\
For a more detailed investigation of the phase diagram for this Hubbard model and its relationship to the momentum space representation, the reader is referred to the work of Kagan and Efremov(\cite{Kag02}).
\section{Experimental tests}
The most likely candidates for the proposed phase are the transition metals Pd, Ni and Nb. Indeed, some indirect evidence for the onset of protonic superfluidity in these systems has been reported \cite{Kar03}. In view of the difficulty in saturating these hydrides in the bulk, the effect may manifest itself primarily at the surface. 
\\
In addition to the normal determinations of superconductivity such as the Meissner effect, the exothermy associated with the pairing phase transition would be quite considerable and should therefore by readily measurable by infra-red or calorimetric techniques. 
\\
It has been pointed out \cite{Legg05} that the proposed proton pairs would possess both electric quadrupole (E2) and  magnetic dipole (M1) moments; the corresponding transitions should be amenable to spectroscopic observation.

\end{document}